# Integration of QoS aspects in the Cloud Computing Research and Selection System

[1]Manar ABOUREZQ and [2]Abdellah IDRISSI

Computer Sciences Laboratory (LRI)
Computer Sciences Department
Faculty of Sciences, Mohammed V University
Rabat, Morocco
[1]manar.abourezq@gmail.com [2]idriab@gmail.com

*Abstract*—Cloud Computing is a business model revolution more than a technological one. It capitalized on various technologies that have proved themselves and reshaped the use of computers by replacing their local use by a centralized one where shared resources are stored and managed by a third-party in a way transparent to end-users. With this new use came new needs and one of them is the need to search through Cloud services and select the ones that meet certain requirements. To address this need, we have developed, in a previous work, the Cloud Service Research and Selection System (CSRSS) which aims to allow Cloud users to search through Cloud services in the database and find the ones that match their requirements. It is based on the Skyline and ELECTRE IS. In this paper, we improve the system by introducing 7 new dimensions related to QoS constraints. Our work's main contribution is conceiving an Agent that uses both the Skyline and an outranking method, called ELECTREIsSkyline, to determine which Cloud services meet better the users' requirements while respecting QoS properties. We programmed and tested this method for a total of 10 dimensions and for 50 000 cloud services. The first results are very promising and show the effectiveness of our approach.

*Keywords*—*Cloud Computing; Cloud Services; Quality of Service; Skyline; Outranking methods; Multi criteria decision; ELECTRE methods; Block-Nested Loops.*

## I. INTRODUCTION

Cloud Computing refers to software, hardware and datacenters offered as a service over a network and remotely accessible via various devices such as computers, PDAs, smart phones, etc. Although it is a rather new computing paradigm that appeared in the last decade, Cloud Computing capitalizes on concepts that have been proven, such as Virtualization [1], Distributed Computing [2], Grid Computing [3], Web Services [4], Service-Oriented Architecture [5], etc.

One of the early definitions of Cloud Computing was proposed by Wang et al. [6], who defined Cloud Computing as "*a set of network enabled services, providing scalable, QoS guaranteed, normally personalized, inexpensive computing platforms on demand, which could be accessed in a simple and pervasive way*".

Another definition based on the concepts Cloud Computing is built on was proposed by Vouk in [7], stating that Cloud Computing "*embraces cyber infrastructure and builds upon decades of research in virtualization, distributed computing, grid computing, utility computing, and, more recently, networking, web and software services. It implies a service-oriented architecture, reduced information technology overhead for the end-user, greater flexibility, reduced total cost of ownership, on-demand services and many other things*".

In [8], Cloud Computing is defined as being a "*type of parallel and distributed system consisting of a collection of interconnected and virtualized computers that are dynamically provisioned and presented as one or more unified computing resources based on service-level agreements established through negotiation between the service provider and consumers*".

The NIST [9] defines Cloud Computing as being "*a model for enabling ubiquitous, convenient, on-demand network access to a shared pool of configurable computing resources that can be rapidly provisioned and released with minimal management effort or service provider interaction*".

Foster et al. propose another definition in [3] where Cloud Computing is considered as "*a large-scale distributed computing paradigm that is driven by economies of scale, in which a pool of abstracted, virtualized, dynamically-scalable, managed computing power, storage, platforms, and services are delivered on demand to external customers over the Internet*".

Though there are many attempted definitions of Cloud Computing, they all agree that every Cloud system has the following essential characteristics:

- The use of virtualization to offer a set of shared physical and virtual resources such as networks, servers, storage space, bandwidth, applications…;

- Dynamic configurability that makes it easy to expand or decrease depending on the user's needs, without affecting the level of reliability and security;

- Accessibility via a network, usually the Internet, from various machines (computers, smart phones, tablets, PDAs…) using standard APIs;

- The use of specific measure systems to control and optimize the use of resources and to offer a billing based on what was consumed, without surplus or need of managing the underlying infrastructure.

The services reachable via Cloud may be divided into three categories [10]: Software as a Service (SaaS), Platform as a Service (PaaS) and Infrastructure as a Service (IaaS). Each one of these categories has specific characteristics that make it more adapted to certain user groups. For instance, enterprises will more likely purchase IaaS and PaaS services, while individuals will be more inclined to use SaaS services.

SaaS [11] allows users to remotely access applications that run in the Cloud's infrastructure by using thin or thick clients. Thus, there is no need to invest in an infrastructure or to buy software licenses. For providers, costs of installation, hosting and maintenance are optimized since many users access to the same application. Examples of SaaS include Google Drive [12] (formerly Google Docs) and Salesforce CRM [13].

PaaS [14] offers a software layer or a development environment as a service on which users will build and deploy their own applications. That way, users won't need to manage the infrastructure while keeping control of the deployed applications and configuring the hosting environment. Examples of PaaS include Salesforce's Force.com [15], Google App Engine [16] and Microsoft Windows Azure [17].

IaaS [18] provides as a service basic storage and computing resources such as servers, network equipments, data warehouses… These resources will be used to run users' own applications. Usually, IaaS satisfies best the end-users' needs of interoperability and portability [19] because they choose the various blocks that compose the infrastructure used. Examples of IaaS include Amazon Elastic Compute Cloud [20] and Microsoft SQL Azure [21].

In addition to these three main models, many others have been proposed such as Hardware as a Service [6], Communication as a Service [22], Network as a Service [23], Data as a Service [22], Workplace as a Service [24], Security as a service [25], Business Process as a service [26], Identity and Policy Management as a Service [27], STorage as a Service [28], Cluster as a Service [29], etc.

Cloud services can be deployed in various models [30], depending on the use case, the provider's business model... The most widespread deployment models are Public, Private, Community and Hybrid.

A Public Cloud [30] is an open Cloud provided by an organization to the general public. It can be accessed via a network, usually the Internet. However, the fact that the Cloud is public doesn't imply that services are offered for free or that the data exchanged by its means is not confidential.

A Private Cloud [30] is offered to the sole use of one organization that either manages it or delegates its management to a third-party. The main advantage of this deployment model is that there are no limitations regarding bandwidth or security, since the resources are exclusively used by the organization.

A Community Cloud [19] is a Cloud shared by organizations belonging to the same community. They can manage their Cloud themselves or delegate the chore to a third-party.

A Hybrid Cloud [31] contains two or more of the Cloud models cited above interconnected by standard or proprietary technologies.

In addition to these four deployment models, new ones are emerging, like the On-Site Private Cloud [19] and the Special Purpose Cloud [32].

The On-Site Private Cloud is a Cloud intended for the private use of a sole organization, just like the Private Cloud. However, it is hosted by the organization, either in a centralized or distributed way. The security aspect is also managed by the organization.

The Special-Purpose Cloud provides, on top of standard resources, additional methods regarding specific use cases. An example that illustrates this model is Google's App Engine with the specific capacities it offers to document management.

Using a Cloud service presents many advantages to end-users. First, there is a significant cost reduction, since users purchase only the resources they need, without surplus or need to invest in infrastructure or maintenance. There's also the guarantee of instant and uninterrupted access to computing and storage resources to any user who has a network connected machine. In addition to it, users can easily adapt the available resources to their specific needs and can add resources as required.

All these advantages have led to an increase in the use of Cloud Computing. With this increase, many new needs have emerged, among which there is the need to find Cloud services that match the users' requirements. Our contribution is in this research area and consists of a Cloud Services Research and Selection System (CSRSS) based on the Skyline and ELECTRE IS as presented in [35] and [36].

The CSRSS allows users to specify the technical and functional requirements of the cloud services they want to use. To do so, it connects to a database of Cloud services and selects the ones that match the users' requirements while giving them the possibility of getting the optimal value of some of these requirements.

With the CSRSS returning the Cloud services that satisfy best the technical and functional requirements specified by users, our objective now is to address Quality of Service (QoS) requirements to better refine the resulting services by keeping the ones that best satisfy QoS parameters.

Our paper is organized as follows. In the next section we present some related work. In section 3, we present the Cloud Service Research and Selection System (CSRSS) as presented in [36]. In section 4, we present QoS aspects and how we integrated them into the CSRS System. In section 5, we present our improved prototype and test results before concluding in section 6.

## II. RELATED WORK

The increase use of Cloud Computing has resulted in the emergence of new needs, such as the need of having systems to search and select Cloud services that meet users' requirements. Many works have been carried out to offer new solutions that will help users to choose the Cloud services that answer their needs. It is rather different from the selection of Cloud service components for composition purpose, which is beyond the scope of our research subject. Our main goal is to find Cloud services that best match the users' requirements, not the selection of two or more Cloud services to compose one final Cloud service that will be delivered to users.

One of the first works dealing with Cloud services discovery and selection was proposed by Goscinski et al. in [29]. It focused on Cloud clusters and used a broker that dynamically matches services and clusters.

Zeng et al. presented a Cloud service selection algorithm in [37]. The algorithm determines the cost and gains of available Cloud services that can be reached via proxy and return those that maximize the gains and minimize the cost. It is done in two steps. In the first step, the proxy selects the available Cloud services following the request sent by the user. In the second step, the algorithm computes the gains and cost of the selected Cloud services and returns the ones that optimize the cost and gains.

Kang and Sim presented a Cloud portal with a Cloud service search engine in [38]. This system uses the concept of similarity [39] and consults the adopted Cloud ontology to select the Cloud services that match the requirements specified by the user.

Kang and Sim also proposed Cloudle in [40], a Cloud services search engine which main functionalities are query processing, similarity reasoning and rating. Like the portal presented in [38], Cloudle consults a Cloud ontology to compute the similarity between Cloud services and returns a list of results sorted by aggregated similarity.

In [41], Han and Sim presented a Cloud Service Discovery System. It consults a Cloud ontology to compute the similarity between Cloud services and return a list of results matching the user's query.

In the three systems presented above, users specify the requirements that must be satisfied by the Cloud services they are searching for. These requirements can be split into three types, namely functional requirements (category of service), technical requirements (OS, CPU, memory, storage space...) and cost requirements (price and timeslot range).

Yoo et al. presented in [42] a resource selection service based on Cloud ontology. It generates Virtual Ontologies (Vons) based on virtualized resources and combine them into new resources. Then it computes the similarity between these new resources to determine the ones that meet best the user's requirements.

In [43], Zang et al. presented a service matching algorithm and a service composition algorithm. These algorithms search through Cloud services and compute the semantic similarity [39] between them to determine whether two given Cloud services are interoperable.

These systems mostly use similarity [39] to determine which Cloud service is the most similar to the user's quest. Thus, they would be better suited for users who want to find Cloud services that are similar to the ones they already know or use. This leaves out users that want to find Cloud services that meet some requirements (service model, provider, bandwidth, latency...) without knowing an existing Cloud services that does meet these requirements. This is why we have thought of replacing similarity with the principle of the Skyline [33].

Using the Skyline allows users to specify the criteria they want to optimize and to get the Cloud services that meet their needs. Thereby, we have developed in [35] a system that enables them to do so and that is based on the principle of the Skyline. We then tried to improve our system by applying outranking methods [34] to the results returned by the Skyline.

There are many works that have used MCDM methods to address the selection of Cloud services. L. Sun et al. conducted a thorough study of Cloud service selection techniques in [44], including MCDM-based techniques such as Analytic Hierarchy Process (AHP) [45], Analytic Network Process (ANP) [46] MAUT [47] and outranking methods [34].

Garg et al. presented a framework for ranking Cloud services based on AHP in [48]. They used the metrics that form the Service Measurement Index (SMI) [49] to define the criteria upon which Cloud services are to be evaluated and compared. They chose six criteria which are accountability, agility, assurance, cost, performance, and security. Although this method is interesting, it has only been tested using three Cloud services as an input in one use case, and 1000 providers in the other.

Godse et al. proposed in [50] an AHP-based approach for the selection of SaaS products Cloud services. The criteria used are functionality, architecture, usability, vendor reputation, and cost. Like the previous one [48], this method has only been tested using three Cloud services (SaaS products). Also, it is only used to compare SaaS Cloud services, leaving out other Cloud services categories.

Karim et al. presented in [51] a QoS mapping approach for combining SaaS and IaaS products and then ranking the combined Cloud services for end users. This method is carried out in four mapping steps. First, the users' QoS requirements are mapped to the QoS specifications of available SaaS products. Second, the obtained SaaS products are mapped to the available IaaS products that have the best QoS guarantees. Third, the end-to-end resulting QoS specifications are computed. Finally, AHP is used to rank the combined Cloud services based on the end-to-end QoS specifications obtained. Tests have been carried using four Cloud services and eight QoS criteria. It is also intended to be used for IaaS and SaaS Cloud services only.

Menzel et al. present in [52] a Multi-Criteria Comparison Method for Cloud Computing, denoted $(MC_2)^2$, that offers a framework for selecting Cloud services using ANP, which is an extension of AHP proposed by Saaty in [46]. This framework allows users to select the best adapted IaaS to their needs. Nine criteria are used such as flexibility to change, reliability, security, maturity of the provider…The $(MC^2)^2$ has been implemented as a web application; AOTEAROA.

Limam and Boutaba proposed in [53] a Cloud service selection approach based on MAUT and aimed at SaaS products. In order to use the MAUT method, the three criteria initially chosen, namely reputation, quality, and cost are reduced to one criterion; feedback.

Silas et al. presented in [54] a middleware for the selection of Cloud services using ELECTRE. Like most of the works cited above, the criteria used are QoS related. The middleware uses ELECTRE III to rank the Cloud services according to the degree to which they match the user's requirements and preferences.

To our knowledge, no work has combined the use of the Skyline operator and ELECTRE. Our motivation to do so in [36] is 1) to capitalize on the results of our first work [35] and refine its results and 2) to minimize the complexity that comes from using ELECTRE alone. Indeed, ELECTRE carries a pairwise comparison to build a decision matrix which size is n x n, n being the number of alternatives. The prior use of the Skyline operator allows making a first filtering of the input, reducing its size up to more than 40%. Thus, the alternatives that are contained in the Skyline form the input to the ELECTRE algorithm, knowing that the Skyline contains all the interesting alternatives for the user, no matter how they weight their preferences. In other words, we apply ELECTRE to less than 60% of the candidate alternatives, after eliminating the rest using the Skyline operator.

We present the prototype and algorithms of our system, as presented in [36], in the next section.

III. THE CLOUD SERVICE RESEARCH AND SELECTION SYSTEM (CSRSS)

The figure below illustrates the prototype of our CSRS System as presented in [36]. It involves the introduction of several agents and consists of a user interface, a user's query processing agent, a pre-Skyline processing agent, a cloud services research and selection agent, an ELECTRE IS agent and a database.

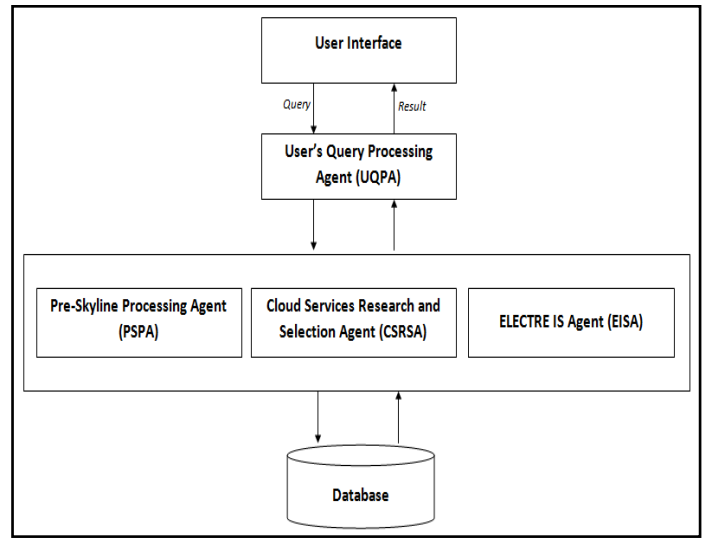

Fig. 1. A schema representing the new version of the Cloud Service Research and Selection System (CSRSS)

The user's interface allows users to interact with the system by selecting the requirements that the Cloud services must meet and view the returned results. It also allows the users to add Cloud services by filling in their attributes such as the name, the provider, the bandwidth, the OS, etc. We think that these requirements are the common ground to existing and upcoming Cloud ontologies [22, 38, 40, 41, 42, 55].

The user's query processing agent extracts the requirements contained in the user's request and sets them into two categories:

- Requirements that are fixed, such as the provider's name, the service model, the OS…;

- Requirements that are to be optimized, such as the price (to be minimized), the bandwidth (to be maximized), etc. These requirements will be used as the Skyline's dimensions.

The Cloud Services Research and Selection Agent (CSRSA) connects to the database and executes a SQL query, which predicates are the fixed requirements returned as a result by the user's query processing, to select all the Cloud services that meet these fixed requirements.

The Pre-Skyline Processing Agent (PSPA) prepares the results extracted from the database by the CSRSA for the running of the Skyline operator. The Cloud services returned and their dimensions are stored as tuples. The dimensions used are the user's requirements that are not "fixed", and thus are to be optimized, such as price (to be minimized), bandwidth (to be maximized), network latency (to be minimized)…

TABLE I. EXAMPLE OF FIXED REQUIREMENTS

| Requirement | Value |
|---|---|
| Provider | Microsoft<br>IBM<br>Amazon… |
| Service Model | IaaS<br>PaaS<br>SaaS |
| OS Serie | Windows<br>Mac<br>Unix… |
| OS Distribution | Windows XP<br>Windows Vista<br>Windows 7<br>Linux… |
| CPU Manufacturer | Intel<br>IBM<br>AMD… |
| CPU Gamme | Pentium<br>Intel 64… |
| Industry | General<br>Education<br>Healthcare… |
| Category | General<br>CRM<br>E-procurement… |

TABLE II. VALUE RANGE OF THE DIMENSIONS USED IN THE SKYLINE

| Dimension | Value range |
|---|---|
| Storage space | 0.14 – 4 000 |
| Memory | 128 – 16 000 |
| Bandwidth | 0 – 10 |
| Latency | 0 – 10 000 |
| Cost | 1 – 2 000 |
| CPU speed | 50 – 3 060 |

The CSRSA uses the Skyline [33], on the set of tuples returned by the PSPA, to determine which Cloud services are in the Skyline and meet the user's preferences.

The Skyline was introduced to meet the needs of users who want to select a set of points that optimize their requirements from a large set of data. Each point contained in the Skyline is not dominated by any other point, thus being better than all the points not contained in the Skyline for at least one criterion, and being equal to or better than them for all the other criteria. A criterion used by the Skyline is called dimension.

For example, a user wants to rent a car at the minimum price with the maximum engine power. In this case, we have two dimensions upon which the selection is to be made: the first dimension is the price; the second is the engine power. The Skyline algorithm will compute the Skyline, which will contain all the cars that are not dominated by any other car. In other words, for each car returned in the Skyline, there is no car outside the Skyline that is less expensive and has more engine power at the same. Thus, a user will find their favorite car in the Skyline, no matter how they weight their preferences toward the dimensions.

In our case, using the Skyline allows the user to specify the criteria they want to optimize and to get the Cloud services that are not dominated by any other Cloud service, that is to say Cloud services for which there exists no better Cloud service for all the criteria specified.

There are two major ways to compute the Skyline. One is to extend existing database systems with the logical Skyline operator. The other is to use algorithms. Many algorithms may be used such as the Block-Nested Loops algorithm (BNL) [33], the Divide and Conquer algorithm (D&C) [56, 57], the B-Tree algorithm [58], etc. We used the BNL algorithm (*Fig.* 2) because it is efficient, simple to implement. It has a complexity of $O(n^2)$ in the worst cases and $O(n)$ in the best.

---

- $L_P$ : input list of tuples for which the Skyline is to be computed
- $L_D$: input list of dimensions
- $p, q$: tuples
- $L_S$ : output list of the tuples forming the Skyline

**Function** *ComputeSkyline*
**Foreach** *p* in $L_P$ **do**

    **If** $L_S = \emptyset$ **Then**

      $L_S = \{p\}$

    **Else**

      **Foreach** *q* in $L_S - \{p\}$ **do**

        result = Compare (p, q, $L_D$)

        **If** result = count ($L_D$) **then**

    $L_S = L_S + \{p\} - \{q\}$

    **Elseif** result # 0 **and** *q is the last tuple in $L_S$* **then**

    $L_S = L_S + \{p\}$

        **Else**

          Goto (*)

    **End IF**

    **End Foreach**

(*) **End If**
**End Foreach**
**Return** $L_S$
**End Function**

---

*Algorithm 1*: Algorithm of the Skyline Agent as presented in [35]

As presented in [36], the BNL algorithm consists of comparing tuples among them to determine the ones that are not dominated by any other. It is done by keeping dominating tuples in the main memory and by comparing each new tuple to them. In each iteration, a new tuple is read from the input list of tuples. If the new tuple is dominated by one of the existing tuples in the main memory, it is eliminated. If it dominates a tuple in the main memory, the dominated tuple is eliminated, and the new tuple is added to the main memory to be compared to future tuples. If the new tuple is incomparable, which means that it is neither dominated by nor dominating any tuple in the main memory, it is added to the main memory.

At the end of all iterations, only tuples that are not dominated by any other tuple are kept in the main memory. These tuples form the Skyline.

The function Compare(p, q, $L_D$) (*Fig. 2*) as presented in [36] compares the tuples p and q in all the dimensions in the list $L_D$. The result returned varies between 0 (when q dominates p) and the number of dimensions n (when p dominates q). Any other result in this range means that p and q are not comparable.

All the tuples contained in the output list are incomparable among them. This means, if we take any given two tuples, each one would be at least better than the other in some dimensions, and at least worse in others. The Skyline doesn't allow arbitrating between incomparable tuples. This comes from the fact that all dimensions are considered to have the same importance, which is not always true to users. To overcome this limitation, we thought of using outranking methods, more specifically ELECTRE methods.

As seen previously, the Cloud Service Research and Selection Agent (noted CSRSA) uses the Skyline on the set of tuples returned by the Pre-Skyline Processing Agent (noted PSPA) to determine which Cloud services are in the Skyline and meet the user's preferences. To do so, the CSRSA uses the BNL algorithm as showed in *Fig. 2*. In order to refine the results returned, we adjusted our prototype (*Fig. 3*) by adding an ELECTRE IS agent. This agent uses the ELECTRE IS algorithm (*Algorithm 2*) to apply the ELECTRE IS to the Skyline list returned by the CSRSA.

---

- p, q: tuples
- c': concordance level
- $L_P$: input list of tuples for which the Skyline is to be computed
- $L_S$: list of the tuples forming the Skyline
- $L_C$: input list of criteria with their information (thresholds...)
- $L_{ES}$: output list of the tuples forming the solution

**Function** ComputeSolution
    $L_{ES} = L_S$
    **Foreach** p in $L_S$ **do**
        **Foreach** q in $L_S$ – {p} **do**
          concordanceIndex = Concordance(p, q, $L_C$)
          vetoIndex = Veto(p, q, $L_C$)
          **If** concordanceIndex ≥ c' **and** vetoIndex = true **then**
              $L_{ES} = L_{ES}$ – {q}
          **End if**
        **End Foreach**
    **End Foreach**
    **Return** $L_{ES}$
**End Function**

---

*Algorithm 2*: Algorithm of the ELECTRE IS Agent as presented in [36]

The Skyline list becomes the input list of alternatives on which the ELECTRE IS method is applied. Thus, a pairwise comparison is made and the concordance and veto indexes are determined. If the validating condition is verified, the alternative that is outranked by the other alternative is deleted from the list of the final solution. Thus, the output list contains only the alternatives that are incomparable both to the Skyline and the ELECTRE IS agents.

In the next section we present Cloud related Quality of Service (QoS) requirements and in particular the ones we use as new dimensions/criteria in the CSRS System.

## IV. QUALITY OF SERVICE (QOS)

As users increasingly turn to Cloud services providers to purchase their services, they are more and more demanding when it comes to Quality of Service (QoS).

QoS is defined as being the "*totality of characteristics of a telecommunications service that bear on its ability to satisfy stated and implied needs of the user of the service*" [59].

There are many important QoS parameters to take into account when looking for a Cloud service, such as time, cost, reliability, security [60]…

The requirements of Cloud users regarding QoS parameters are described in Service-Level Agreements (SLAs) to help providers manage the services contracted and maintain the overall level of quality agreed on with end-users [61]. And since Cloud resources are consumed simultaneously by multiple users/tenants, providers have to dynamically allocate Cloud resources among them while guaranteeing the QoS level agreed on for every one of them. So for Cloud users, QoS and SLA are key factors when they select Cloud services.

Measuring the performance of Cloud services is not an obvious task. For one part, there is the question of quantifying parameters that are essentially qualitative. Many works have tried to provide a set of QoS parameters that can be used by Cloud users to select the most adapted services.

In [62], Cao et al. present a QoS-assured Cloud Computing architecture to answer QoS-related requests from users. This architecture consists of X layers: physical device and virtual resource, cloud service provision, cloud service management, and multi-agent. The QoS attributes considered are related both to the users and the providers of cloud services. Many attributes were defined, namely response time, cost, availability, reliability and reputation.

In [63], Ferreti et al. propose a middleware architecture to configure, manage and optimize cloud services in accordance with users' QoS requirements such as timeliness, scalability, availability, and security. The proposed architecture integrates three main components, namely dynamic resources configuration, platform monitoring, dispatching and load balancing of requests and resources. The cloud computing environment resulting is labeled "QoS-aware".

Bouguettaya et al. presented in [64] a QoS-based approach for the selection of cloud services for composition purposes. The aim of this approach is to compose a cloud service that answers the QoS requirements of end-users from multiple composite services provided by different providers. It is done by constructing the composition schema based on the user's request, then selecting the optimal composition plan based on the end-user's QoS requirements. The QoS attributes used are throughput, response time, and cost.

In [65], Zheng et al. presented CloudRank, a QoS ranking prediction framework for cloud services that takes into account users' experiences. In this work, QoS attributes are divided in two categories: client-side and server-side. The latter include response time, throughput, failure probability, etc. and are the ones used in CloudRank. It also uses similarity to determine the degree to which the current user is similar to other users in order to predict which Cloud services would be more interesting.

In [66], Nathuji et al. developed Q-Clouds, a control framework that supports QoS-aware cloud environments, as presented in [63]. Q-Clouds adapts the allocation of resources to absorb the effect of performance interferences that are bound to happen, since many users share the same resources. This is done while taking into account the QoS requirements of users.

In [67], Serrano et al. address the challenge of QoS and SLAs management in Cloud environment by defining a new Cloud model called SLA Aware Service (SLAaaS) and a new language to describe QoS-oriented cloud SLAs, called CSLA, that is inspired from WSLA (SLA for web services) [68] and SLA for Service Oriented Architecture (SLA@SOI) [69].

CSLA formalizes the SLA between users and providers by translating QoS requirements into clauses combined using Boolean operators. QoS attributes adopted in this work are related to performance, availability, reliability and cost.

Another prominent work is the Service Measurement Index (SMI) developed by the Cloud Service Measurement Initiative Consortium (CSMIC) [49].

Service Measurement Index (SMI)is defined by the CSMIC as being "*a set of business-relevant Key Performance Indicators (KPI's) that provide a standardized method for measuring and comparing a business service regardless of whether that service is internally provided or sourced from an outside company*".

SMI measures performance using six categories (*Fig. 4*): agility, risk, security, cost, quality, and capability. Each category contains many attributes.

- Agility: one of the main reasons why organizations choose to move to the Cloud is to increase their agility in order to quickly adapt to their ever-changing business environment. In order to measure agility, SMI proposed a set of parameters that show how quickly and efficiently providers integrate new capabilities to answer users' evolving needs. We retained from these parameters portability.

- Risk: risk is an inherent in IT, the main objective of organizations being not to annihilate it, but to minimize its causes and effects. The main issue for users when choosing a Cloud service provider is verifying the reputation of the latter and making sure they have an efficient risk management strategy. We chose to retain as a parameter the number of risk management certifications the provider has.

- Security: moving to the Cloud can be challenging for users when it comes to entrusting their critical data to providers. Many aspects must be addressed, mainly privacy and data loss. This issue is also related to the laws governing the geographical location of data storage. We chose to initially retain data loss as a parameter.

- Cost: another main reason why organizations and users move to the Cloud is to optimize their IT-related costs. With the Cloud being a pay-as-you-go utility, users are guaranteed to pay solely for resources they consume, without surplus or need to invest or manage the underlying infrastructure. We chose to retain the two cost parameters defined by SMI, namely the on-going cost, which is the cost regularly paid by the Cloud tenant in exchange for the resources they use, and the acquisition cost, which is the cost of the changes necessary to move to the Cloud.

- Quality: when choosing to move to the Cloud, users usually worry about the quality of the services offered by providers, especially as regards their reliability and availability. This comes from the fact that users will be moving their data out of their control and into that of the provider. We chose to retain as parameters availability and service response time.

- Capability: SMI proposes to measure the overall ability of a cloud services provider to satisfy users' requirements by comparing the services offered to standards. We refrained from using a parameter to translate this characteristic since, in our knowledge, there are no unified Cloud standards yet.

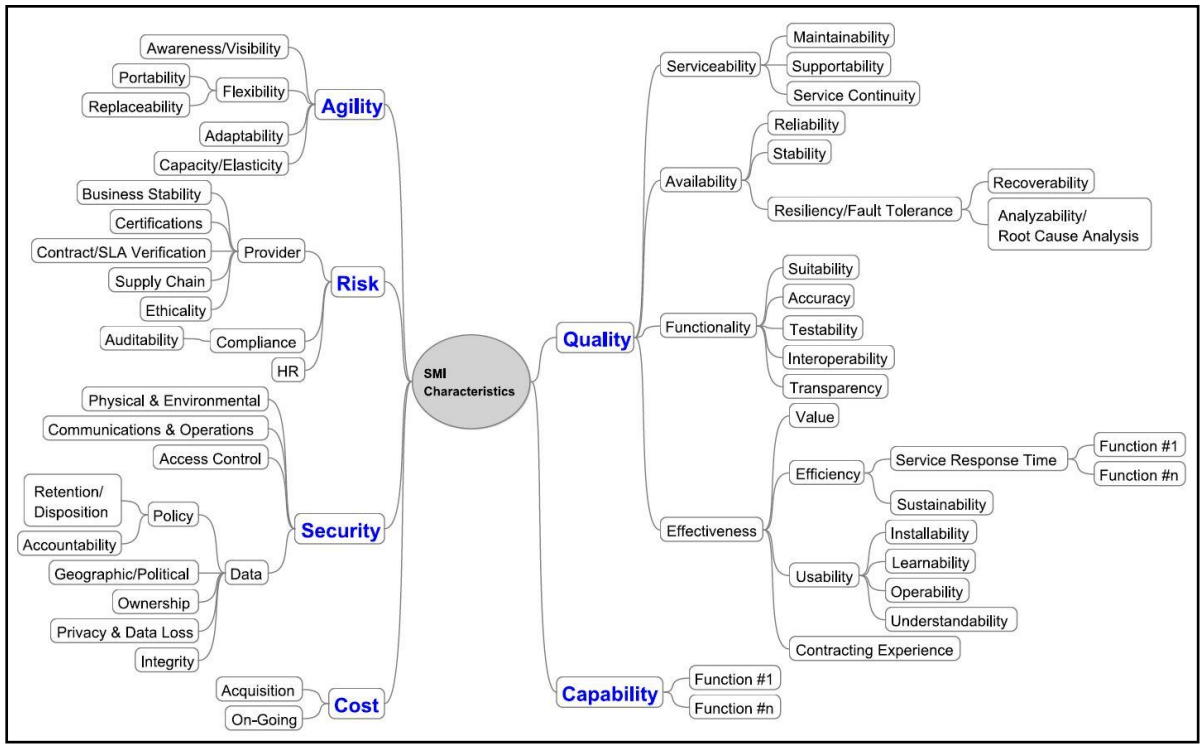

Fig. 2. Service Measurement Index (SMI) characteristics [49]

As explained above, we have chosen to use one or two parameters in each category. These parameters will be used as additional dimensions for the Skyline and criteria for ELECTRE IS. For each new dimension/criterion, we propose a measurement method, as presented in *Table III* hereafter.

TABLE III. QOS PARAMETERS USED AS DIMENSIONS FOR THE SKYLINE

| Category | Dimension/Criterion | Detail | Comparison sense |
|---|---|---|---|
| Agility | Portability | $\frac{\text{number of OS compatible with the service}}{\text{number of OS required by the user}}$ | Maximize |
| Risk | Number of risk management certifications obtained by the provider | – | Maximize |
| Security | Data Loss | Number of data loss related incidents | Minimize |
| Cost | Acquisition cost | – | Minimize |
| Cost | On-going cost | – | Minimize |
| Quality | Service response time | $\frac{\text{average response time (ms)}}{\text{maximum Sresponse time defined in the SLA (ms)}}$ | Minimize |
| Quality | Availability | $\frac{\text{time during which the service is unavailable (ms)}}{\text{total time of use (ms)}}$ | Minimize |

In the next section we present the implementation of the algorithm and its performance along with some screenshots illustrating its execution.

V. EXPERIMENTATION AND RESULTS

The platform we used for the experiments [35] is an HP workstation with a 3.30 GHz processor, 4 GB of main memory, Windows Server 2008 as operating system and MS SQL Server 2008 as DBMS. The algorithm is implemented using ASP.net to obtain a web-based system that can be accessed from any web client anytime the user is connected to the Internet.

The CSRSS start page (Fig. 3) allows the user to either add a new Cloud service to the database or search for Cloud services that match their requirements.

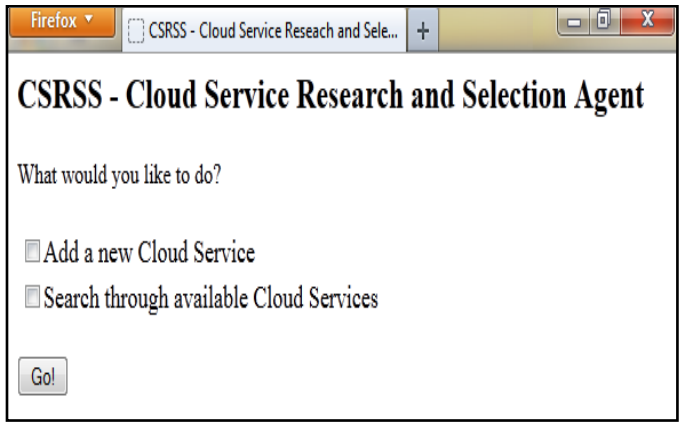

Fig. 3. The CSRA start page

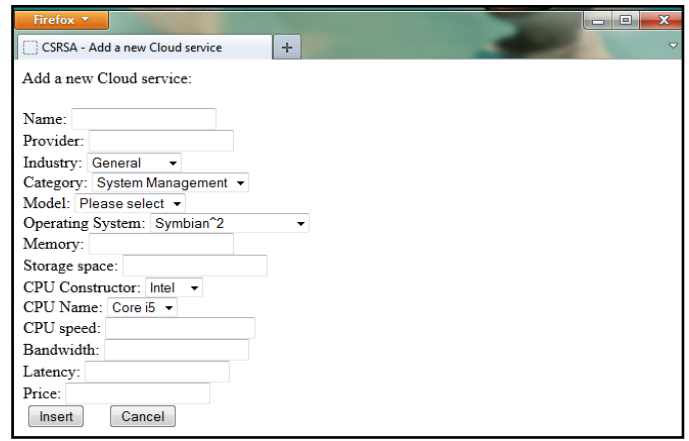

Fig. 4. The CSRA page to add a new Cloud service

If the user chooses to add a new Cloud Service, they are taken to another page (*Fig. 4*) where they first enter the name of the Cloud service in question so a search can be made to make sure that it doesn't already exist in the database. Afterwards, the user enters the different information such as the Cloud service's provider, model (IaaS, PaaS or SaaS), industry, memory, price…

If the user checks the second option (Search through available Cloud Services), they are taken to the CSRSS page (*Fig. 5*) that allows to make an advanced search through the database and to run the algorithm on the returned result in order to obtain the final refined set of Cloud services.

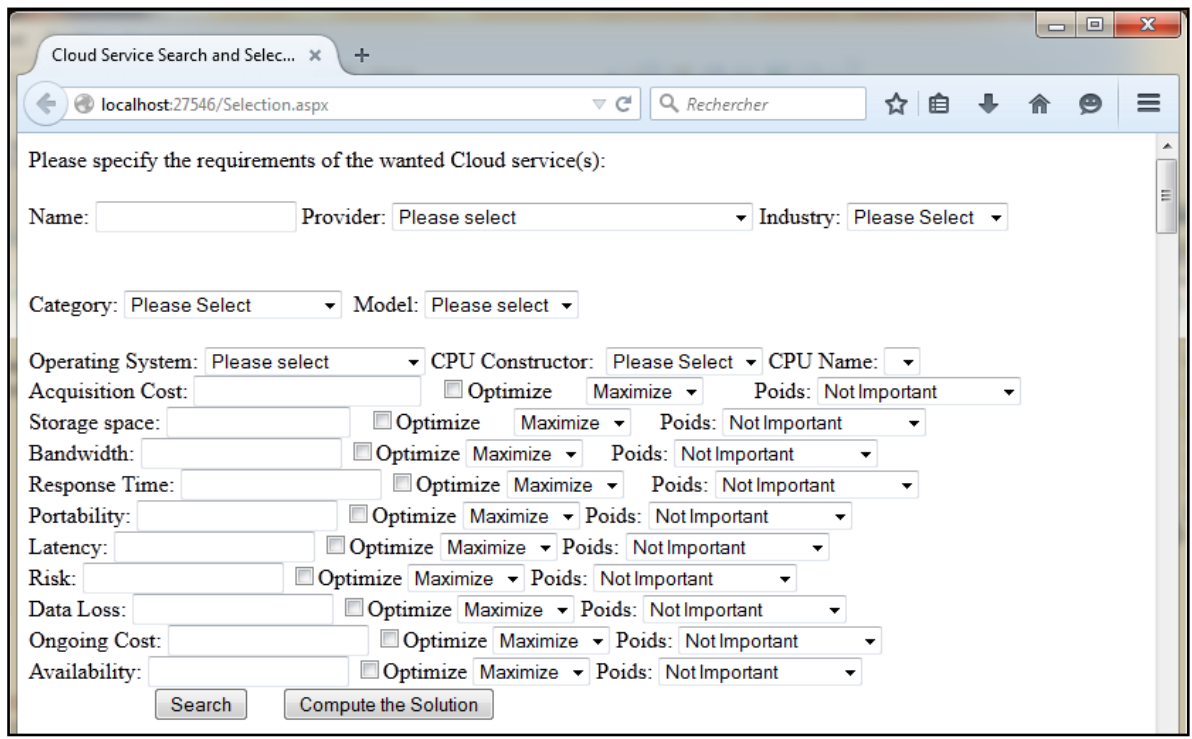

Fig. 5. The CSRA search and/or computation of the Solution page

The user can fill out one or many information about the Cloud service(s) they are searching for. For information such as price, memory, storage space, bandwidth... they can either give a specific value or specify that they are the dimensions to be used when computing the Skyline. For each dimension, the user specifies if it is to be minimized or maximized. They also specify its importance (on a scale from "not important" to "extremely important") that is to be translated into a weight in order to use the ELECTRE IS method.

We worked on the same generated data as in [35]. This data consists of 50 000 Cloud services which we completed with the new 7 dimensions described in Table III. We also chose to disregard dimensions that are too oriented for a specific Cloud service model, such as RAM and CPU speed, which are mostly relevant in IaaS environments, for instance. Thus, we obtained a total 10 dimensions/criteria for each cloud service. Dimensions' values are randomly generated within the ranges specified in Table IV hereafter.

TABLE IV. VALUE RANGE OF THE DIMENSIONS USED IN THE SKYLINE

| Dimension | Value range | Comparison sense |
|---|---|---|
| Storage space | 0.14 – 4 000 | Maximize |
| Bandwidth | 0 – 10 | Maximize |
| Latency | 0 – 10000 | Minimize |
| Portability | 0.03 – 400 | Maximize |
| Risk | 0 – 400 | Maximize |
| Data Loss | 0 – 9 000 | Minimize |
| Acquisition cost | 1 – 20 000 | Minimize |
| On-going cost | 0.1 – 2000 | Minimize |
| Service response time | 0 – 40 | Minimize |
| Availability | 0 – 1 000 | Minimize |

We executed our program varying the number of dimensions from 1 to10 and the input size from 100 to 50 000 cloud services (*Fig. 6*).

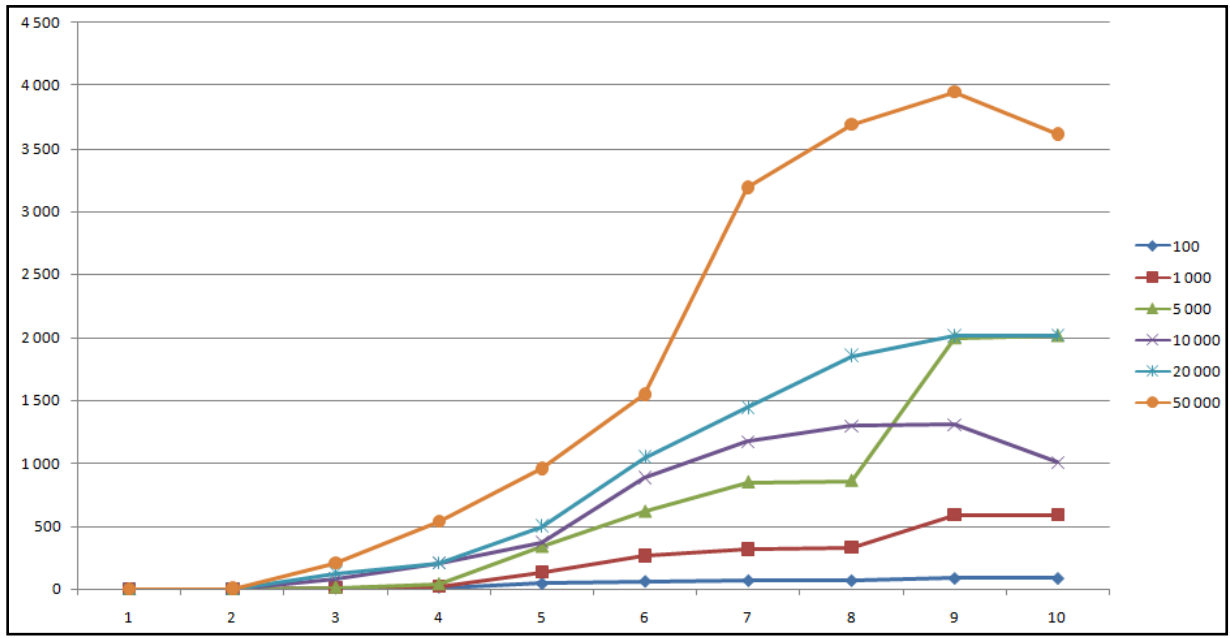

Fig. 6. The size of the solution depending on the number of dimensions and the input size for the ELECTREIsSkyline Algorithm

We also compared the results obtained using ELECTREIsSkyline algorithm with those of using the Skyline algorithm as presented in [35]. We did so for an input list consisting of 50 000 Cloud services and varying the number of dimensions from 1 to 10 (Table V and *Fig. 7*).

TABLE V. SIZE OF THE FINAL SOLUTION DEPENDING ON THE NUMBER OF CRITERIA AND THE ALGORITHM USED FOR 50 000 CLOUD SERVICES

| Number of criteria | Final solution's size | |
|---|---|---|
| | Skyline algorithm | ELECTREIsSkyline algorithm |
| 1 | 18 | 1 |
| 2 | 25 | 7 |
| 3 | 1436 | 205 |
| 4 | 3111 | 539 |
| 5 | 3448 | 957 |
| 6 | 4316 | 1 546 |
| 7 | 6918 | 3 187 |
| 8 | 5285 | 3 688 |
| 9 | 5286 | 3 945 |
| 10 | 7 360 | 3 610 |

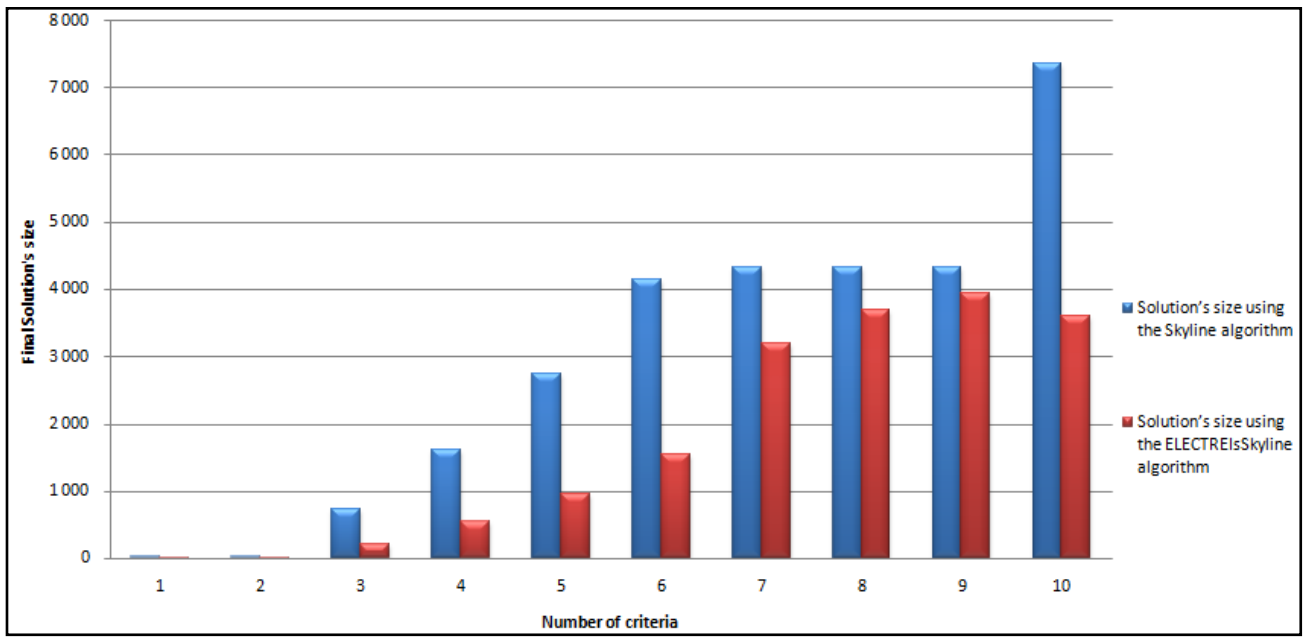

Fig. 7. The size of the solution depending on the number of dimensions for each used algorithm

The use of the ELECTREIsSkyline algorithm proves to be more efficient in determining the cloud services that best match the users' requirements, including QoS requirements. It is due to the fact that users' preferences towards each criterion are taken into account. Thus, dimensions that had the same weight to the Skyline algorithm have different weights in ELECTREIsSkyline, depending on their importance in the decision making process. So, cloud services that were incomparable when using the Skyline become comparable when using ELECTREIsSkyline. The size of the final solution can be reduced to contain only 6% of the input size, returning a total of 3 000 Cloud services from the 50 000 in the input list, while taking into consideration all 10 criteria and their respective weights. However, the final result is highly dependent on the values of the dimensions/criteria in the input size, and the weights attributed by users to these criteria. Also, it is common that when augmenting the input size, one or more new cloud services prove to be highly efficient and allow eliminating many others, reducing significantly the size of the output. These cloud services are called "killer tuples". On the other hand, it is also common that new cloud services turn out to be incomparable with those contained in the solution list, which contributes to augmenting its size. Thus, the final solution's size is highly depending on the quality of input data and the user's preferences.

## VI. CONCLUSION

With many Cloud providers offering their services, Cloud users may be at loss when wanting to choose an adapted cloud service to their needs. To address this issue, we have presented in [35] and [36] the Cloud Service Research and Selection System (CSRSS) that allows users to select cloud services that best suit them by specifying the requirements they are looking for. In this work, we tried to ameliorate the CSRSS by addressing the issue of QoS and adapting it to integrate QoS parameters, giving users the possibility to specify the values of the QoS attributes they require.

REFERENCES

[1] S. Casselman, "Virtual computing and the virtual computer", FPGAs for Custom Computing Machines, Proceedings IEEE Workshop, 1993
[2] P.H ENSLOW, "What is a "distributed" data processing system?", Computer, Vol. 11, No 1, 1978
[3] I. Foster, Y. Zhao, I. Raicu and S. Lu, "Cloud Computing and grid computing 360-degree compared", Grid Computing Environments Workshop, GCE'08, November 2008
[4] D. Roman et al., "Web service modeling ontology", Applied ontology, Vol. 1, No 1, 2005
[5] M. BELL, "Introduction to Service-oriented Modeling", Service-oriented Modeling: Service Analysis, Design, and Architecture. Wiley & Sons, Vol. 3, 2008
[6] Wang, Lizhe, et al. "Scientific Cloud Computing: Early Definition and Experience", HPCC, Vol. 8, 2008
[7] A. Vouk, "Cloud Computing–issues, research and implementations", Journal of Computing and Information Technology, Vol. 16, No 4, 20
[8] R. Buyya, C.S. Yeo and S. Venugopal, "Market-oriented Cloud Computing: Vision, hype, and reality for delivering it services as computing utilities", High Performance Computing and Communications, 2008, HPCC'08, 10th IEEE International Conference, Ieee, 2008
[9] P. Mell and T. Grance, "The NIST definition of Cloud Computing", NIST special publication, 2011
[10] A.Fox, G. Rean, A. Joseph, R. Katz, A. Konwinski, G. Lee, D. Patterson, A. Rabkin, and I. Stoica, "Above the clouds: A Berkeley view of Cloud Computing", Dept. Electrical Eng. and Comput. Sciences, University of California, Berkeley, Rep. UCB/EECS 28, 2009
[11] L. Vaquero, L. Rodero-Merino, J. Caceres and M. Lindner, "A Break in the Clouds: Towards a Cloud Definition", ACM SIGCOMM Computer Communication Review, Vol. 39, Number 1, January 2009
[12] Google Drive, https://drive.google.com
[13] Salesforce, http://www.salesforce.com
[14] D. Cheng, "PaaS-onomics: A CIO's Guide to using Platform-as-a-Service to Lower Costs of Application Initiatives While Improving the Business Value of IT", Tech. rep., LongJump, 2008

[15] Force, http://www.force.com

[16] Google App Engine, https://appengine.google.com

[17] Windows Azure, http://www.windowsazure.com

[18] L. Karadsheh, "Applying security policies and service level agreement to IaaS service model to enhance security and transition", Computers & Security, Vol. 31, Issue 3, May 2012, pp. 315-326

[19] S. Radack, "Cloud Computing: A Review of Features, Benefits, and Risks, and Recommendations for Secure, Efficient Implementations", NIST, ITL Bulletin, June 2012

[20] Amazon, http://aws.amazon.com/fr/ec2

[21] Microsoft SQL Azure, http://www.windowsazure.com

[22] L. Youseff, L. Butrico and M. Da Silva, "Toward a Unified Ontology of Cloud Computing", Grid Computing Environments Workshop, November 2008

[23] P. Costa et al. "NaaS: Network-as-a-Service in the Cloud" Proceedings of the 2nd USENIX conference on Hot Topics in Management of Internet, Cloud, and Enterprise Networks and Services, Vol. 12, 2012

[24] E. B. Dudin and G. Yu Smetanin, "A review of Cloud Computing", Scientific and Technical Information Processing, Vol. 38, No 4, 2011

[25] M. Christodorescu, R. Sailer, D. L. Schales, D. Sgandurra and D. Zamboni, "Cloud security is not (just) virtualization security: a short paper", Proceedings of the 2009 ACM workshop on Cloud Computing security, ACM, 2009

[26] R. Accorsi, "Business process as a service: Chances for remote auditing", Computer Software and Applications Conference Workshops (COMPSACW), 2011 IEEE 35th Annual, IEEE, 2011

[27] M. Zhou, R. Zhang, D. Zeng and W. Qian, "Services in the Cloud Computing era: A survey", Universal Communication Symposium (IUCS), 2010 4th International IEEE, 2010

[28] J. Wu, L. Ping, X. Ge, Y. Wang, and J. Fu, "Cloud storage as the infrastructure of Cloud Computing", Intelligent Computing and Cognitive Informatics (ICICCI), 2010 International Conference IEEE, 2010

[29] M. Brock and A. Goscinski, "Toward Ease of Discovery, Selection and Use of Clusters within a Cloud," 2010 IEEE 3rd International Conference on Cloud Computing, July 2010

[30] S. Rao, N. Rao and E. Kusuma Kumari, "Cloud Computing: An Overview", Journal of Theoretical and Applied Information Technology, Vol. 9, No. 1, November 2009

[31] K. Sims, "IBM Blue Cloud Initiative Advances Enterprise Cloud Computing", 2009

[32] K. Jeffery and B. Neidecker-Lutz, "The future of Cloud Computing", European Commission, Information Society and Media, 2010

[33] S. Börzsönyi, D. Kossmann, and K. Stocker, "The Skyline operator", International Conference on Data Engineering (ICDE), 2001

[34] B. Roy, "The outranking approach and the foundation of the ELECTRE methods", Theory and decision, Vol. 31, Issue 1, 1991

[35] A. Idrissi and M. Abourezq, "Skyline in Cloud Computing", Journal of Theoretical and Applied Information Technology, Vol. 60, No. 3, February 2014

[36] A. Idrissi and M. Abourezq, "Introduction of an outranking method in the Cloud computing research and Selection System based on the Skyline", Proceedings of the International Conference on Research Challenges in Information Science (RCIS), 2014

[37] W. Zeng, Y. Zhao and J. Zeng, "Cloud Service and Service Selection Algorithm Research", of the first ACM/SIGEVO Summit on Genetic and Evolutionary Computation

[38] J. Kang and K. M. Sim, "A Cloud Portal with a Cloud Service Search Engine", International Conference on Information and Intelligent Computing IPCSIT, Vol.18, 2011

[39] P. Resnik, "Semantic similarity in a taxonomy: an information-based measure and its application to problem of ambiguity in natural language", Journal of Artificial Intelligence Research, Vol. 11, 1999

[40] J. Kang and K. M. Sim, "Cloudle : An Agent-based Cloud Search Engine that Consults a Cloud Ontology", Cloud Computing and Virtualization Conference, 2010

[41] T. Han and K. M. Sim, "An Ontology-enhanced Cloud Service Discovery System", IMECS 2010 Vol. 1, March 17 – 19 2010, Hong Kong

[42] H. Yoo, C. Hur, S. Kim, and Y. Kim, "An Ontology-based Resource Selection Service on Science Cloud", International Journal of Grid and Distributed Computing, Vol. 2, No. 4, December 2009

[43] C. Zeng, X. Guo, W. Ou and D. Han, "Cloud Computing Service Composition and Search Based on Semantic", Cloud Computing, Vol. 5931, 2009, pp. 290-300

[44] L. Sun, H. Dong, F. K. Hussain, O. K. Hussain, and E. Chang, "Cloud service selection: State-of-the-art and future research directions." Journal of Network and Computer Applications 45, 2014

[45] T.L. Saaty, "The Analytic Hierarchy Process for Decision in a Complex World", Pittsburgh, PA: RWS Publications, 1980

[46] T. L. Saaty, "Decisions with the analytic network process (ANP)" University of Pittsburgh (USA), ISAHP 96 (1996)

[47] C. W. Churchman, R. L. Ackoff and E.L. Arnoff, "Introduction to Operations Research", New York: Wiley, 1957

[48] S. K. Garg, S. Versteeg and R. Buyyaa, "SMICloud: A framework for ranking of Cloud Computing services", IEEE International Conference on Utility and Cloud Computing, 2011

[49] Cloud Service Measurement Index Consortium (CSMIC), SMI framework, http//cloudcommons.com/servicemeasurementindex

[50] M. Godse and S. Mulik, "An approach for selecting software-as-a-service (SaaS) product", Proceedings of the IEEE international conference on Cloud Computing (CLOUD), Bangalore, 2009

[51] R. Karim, C. Ding and A. Miri, "An end-to-end QoS mapping approach for Cloud service selection", Proceedings of the IEEE 9th world congress on services (SER-VICES), Santa Clara Marriott, CA, 2013

[52] M. Menzel, M. Schönherr and S. Tai, "(MC2)2: criteria, requirements and a software prototype for Cloud infrastructure decisions", Softw Pract Exp November 2013, Vol. 43, Issue 11

[53] N. Limam and R. Boutaba, "Assessing software service quality and trustworthiness at selection time", IEEE Trans Softw Eng 2010, Vol. 36, Issue 4

[54] S. Silas, E. B. Rajsingh and K. Ezra, "Efficient service selection middleware using ELECTRE methodology for Cloud environments", Information Technology Journal, Vol. 11, Issue 7

[55] D. Androcec, N. Vrcek and J. Seva, "Cloud Computing Ontologies: A Systematic Review", MOPAS 2012, The Third International Conference on Models and Ontology-based Design of Protocols, Architectures and Services, 2012

[56] H. Kung, F. Luccio and F. Preparata, "On finding the maxima of a set of vectors", Journal of the ACM, Vol. 22, Issue 4, October 1975

[57] F. Preparata and M. Shamos, "Computational Geometry: An Introduction", Springer-Verlag, New York, 1985

[58] D. Comer, "The Ubiquitous B-Tree", ACM Computing Surveys, Volume 11, June 1979

[59] "Terms and definitions related to quality of service and network performance including dependability", International Telecommunication Union Recommendation, 1994

[60] R. Buyya, C. S. Yeo and S. Venugopal, "Market-Oriented Cloud Computing: Vision, Hype, and Reality for Delivering IT Services as Computing Utilities", High Performance Computing and Communications, 2008, HPCC'08, 10th IEEE International Conference

[61] M. E. Shacklett, "Five Key Points for Every SLA", http://content.dell.com/us/en/enterprise/d/large-business/key-points-for-sla.aspx, 2011

[62] B. Q. Cao, B. Li and Q. M. Xia, "A Service-Oriented QoS-Assured and Multi-AgentCloud Computing Architecture", Cloud Computing, Springer Berlin Heidelberg

[63] S. Ferretti, V. Ghini, F. Panzieri, M. Pellegrini and E. Turrini, "QoS–aware Clouds", Cloud Computing (CLOUD), 2010 IEEE 3rd International Conference

[64] Z. Ye, A. Bouguettaya and X. Zhou, "QoS-Aware Cloud Service Composition Based on Economic Models", Service-Oriented Computing, Springer Berlin Heidelberg

[65] Z. Zheng, X. Wu, Y. Zhang, M. R. Lyu and J. Wang, "QoS Ranking Prediction for Cloud Services", Parallel and Distributed Systems, IEEE Transactions, Vol. 24, Issue 6

[66] R. Nathuji, A. Kansal and A. Ghaffarkhah, "Q-Clouds: Managing Performance Interference Effects for QoS-Aware Clouds", Proceedings of the 5th European conference on Computer systems, ACM, 2010

[67] D. Serrano, S. Bouchenak, Y. Kouki, T. Ledoux, J. Lejeune, J. Sopena and P. Sens, "Towards QoS-Oriented SLA Guarantees for Online Cloud Services", Cluster, Cloud and Grid Computing (CCGrid), 2013 13th IEEE/ACM International Symposium

[68] H. Ludwig, A. Keller, A. Dan, R. P. King, and R. Franck, "Web Service Level Agreement (WSLA) Language Specification," IBM, Tech. Rep., 2003

[69] "Sla@soi," sla-at-soi.eu/, 2012.